# MEMPSEP III. A machine learning-oriented multivariate data set for forecasting the Occurrence and Properties of Solar Energetic Particle Events using a Multivariate Ensemble Approach

**Kimberly Moreland[1,2], Maher Dayeh[2,1], Hazel M. Bain[4,5], Subhamoy Chatterjee[3], Andrés Muñoz-Jaramillo[3], Samuel Hart[1,2]**

[1]The University of Texas at San Antonio, San Antonio, TX, USA [2]Southwest Research Institute, San Antonio, TX, USA [3]Southwest Research Institute, Boulder, CO, USA
[4]Cooperative Institute for Research in Environmental Sciences, University of Boulder, CO, USA
[5]Space Weather Prediction Center, NOAA, Boulder, CO, USA

**Key Points:**

- Machine Learning oriented dataset for SEP event prediction and subsequent properties
- Multivariate remote sensing and in-situ observations
- Continuous dataset spanning several decades

Corresponding author: Kimberly Moreland, kim.moreland@contractor.swri.org

**Abstract**

We introduce a new multivariate data set that utilizes multiple spacecraft collecting in- situ and remote sensing heliospheric measurements shown to be linked to physical processes responsible for generating solar energetic particles (SEPs). Using the Geostationary Operational Environmental Satellites (GOES) flare event list from Solar Cycle (SC) 23 and part of SC 24 (1998-2013), we identify 252 solar events (flares) that produce SEPs and 17,542 events that do not. For each identified event, we acquire the local plasma properties at 1 au, such as energetic proton and electron data, upstream solar wind conditions, and the interplanetary magnetic field vector quantities using various instruments onboard GOES and the Advanced Composition Explorer (ACE) spacecraft. We also collect remote sensing data from instruments onboard the Solar Dynamic Observatory (SDO), Solar and Heliospheric Observatory (SoHO), and the *Wind* solar radio instrument WAVES. The data set is designed to allow for variations of the inputs and feature sets for machine learning (ML) in heliophysics and has a specific purpose for forecasting the occurrence of SEP events and their subsequent properties. This paper describes a dataset created from multiple publicly available observation sources that is validated, cleaned, and carefully curated for our machine-learning pipeline. The dataset has been used to drive the newly-developed Multivariate Ensemble of Models for Probabilistic Forecast of Solar Energetic Particles (MEMPSEP; see MEMPSEP I (Chatterjee et al., 2023) and MEMPSEP II (Dayeh et al., 2023) for associated papers).

## Plain Language Summary

We present a new dataset that uses observations from multiple spacecraft observing the Sun and the interplanetary space around it. This data is connected to the processes that create solar energetic particles (SEPs). SEP events pose threats to both astronauts and equipment in space. The dataset contains 252 solar events that caused SEPs and 17,542 that do not. For each event, we gather information about the local space environment around the sun, such as energetic protons and electrons, the conditions of the solar wind, the magnetic field, and remote solar imaging data. We use instruments from NOAA's Geo-stationary Operational Environmental Satellites (GOES) and the Advanced Composition Explorer (ACE) spacecraft, as well as data from the Solar Dynamic Observatory (SDO), the Solar and Heliospheric Observatory (SoHO), and the Wind solar radio instrument WAVES. This data set is designed to be used in machine learning, with a focus on predicting the occurrence and properties of SEP events. We detail each observation obtained from publicly available sources, and the data treatment processes used to validate the reliability and usefulness for machine-learning applications.

## 1 Introduction

Solar energetic particles are high-energy particles associated with two main types of solar activity: solar flares (SFs) and coronal mass ejections (CMEs) (M. Desai & Giacalone, 2016; Reames, 2013). SEP flux enhancements last from tens of minutes to days and include protons, electrons, and heavier ions. The proton particle energies can sometimes reach giga electron volts (GeV) (Reames, 2001). Particles of this energy range can negatively affect technological assets in space, (Horne et al., 2013; Maurer et al., 2017) cause high-dose radiation exposure of astronauts and even affect passengers and crews on polar route commercial airline flights (Onorato et al., 2020; Chancellor et al., 2014).

Solar flares originate in the lower solar corona and chromosphere in regions of complex magnetic fields, also known as active regions (ARs). Flares are easily recognized by their large enhancements in the Extreme Ultraviolet (EUV) and X-ray frequency bands. X-rays occur due to collisions between decelerating closed-loop particles and the underlying plasma (Galloway, R. K. et al., 2010). The Solar Dynamics Observatory observes solar flares at eleven passbands of differing wavelengths. The 171 Å band allows for investigating the flare intensity, location, and evolution. Some X-ray flare events have Type III radio bursts (i.e., Figure 3) that are observed as quickly sweeping bursts from high to low frequencies in radio spectrograms. Type III radio bursts are radio waves associated with electrons accelerated at solar magnetic reconnection sites at or near the flaring region that travel along open magnetic field lines through the upper corona and into interplanetary space (Cairns et al., 2018). In addition to flares, active regions can also produce Coronal Mass Ejections (CMEs).

Some ARs develop a build-up of mass along a closed magnetic field line that eventually erupts, releasing a CME. CMEs appear as expanding loops or bubbles and are often seen in visible light coronagraphs (e.g. Large Angle and Spectromic Coronagraph (LASCO)). LASCO records white light images of the solar corona from 1.1 up to 30 solar radii and spectral images of the solar corona from 1.1 to 3.0 solar radii. Coronagraph images make it possible to view the field topology of the corona and its changes, as well as the evolution of CMEs as they travel outward from the sun. CMEs, moving much faster than the preceding material, act as the driver of interplanetary (IP) shocks. One indicator of a shock occurrence is a Type II radio burst (E. W. Cliver et al., 1986; Makelä et al., 2011), which is generally observed as a slower high to low-frequency sweep and is associated with electrons accelerated by the outward propagating shocks. At the time of the shock passage, the local particle flux may be enhanced from suprathermal energies to tens of MeV/nucleon. This sudden, short-lived enhancement is known as an energetic storm particle (ESP) event and is only seen at lower energies (M. Desai & Giacalone, 2016; Moreland et al., 2023). The primary candidate for ESP acceleration at shocks is the diffusive shock acceleration (DSA) mechanism, comprising the shock-drift mechanism at quasi-perpendicular shocks (Decker, 1981) and the first-order Fermi mechanism at quasi-parallel shocks (Lee, 1983). Ideally, DSA theory provides an explanation for several aspects of ESP observations and predicts certain particle profile behaviors (e.g., spectral index). However, numerous issues affect this, including particle transport, localized turbulence, magnetic connectivity along the shock, shock geometry, and other factors (Giacalone & Neugebauer, 2008; Zank et al., 2010; Mostafavi et al., 2018).

Understanding the physical processes behind SEPs and predicting SEP events as well as their properties (i.e. peak flux, energies, onset, duration, etc.) using true probabilities is becoming exceedingly important. Fortunately, the heliophysics and space weather communities now benefit from large amounts of free and publicly available remote sensing and in-situ observations collected over decades. To a greater extent, we can take advantage of the significant advances in computing power, open-source software, and validated algorithms that make for a perfect combination for ML applications (Camporeale, 2019). There are a variety of available datasets in the community such as SHARPs and SMARPs (Bobra et al., 2021), which are derived from MDI and SoHO solar surface magnetic field maps to provide a seamless set of maps and keywords describing each active region observed since 1996; a multivariate time series (MVTS) dataset available via a web API, extracted from SHARPs data and cross-checked with the NOAA solar flare catalog that includes fifty-one solar flare predictive parameters spanning from 2010-2018 ((Angryk et al., 2020)); and a multitude of SEP event lists (e.g. (Kahler et al., 2017; Papaioannou et al., 2016; Crosby et al., 2015; Gopalswamy et al., 2014)). As of yet, none have incorporated a diverse set of multivariate observations that includes remote sensing and in-situ observations at 1 au, which is the purpose of this work. This dataset has also been used to drive a newly developed model for forecasting the occurrence and properties of SEPs, namely, the "Multivariate Ensemble of Models for Probabilistic Forecast of Solar Energetic Particles" (MEMPSEP; see MEMPSEP I (Chatterjee et al., 2023) and MEMPSEP II (Dayeh et al., 2023) associated papers.

When assembling this dataset, we consider each physical process independently and select observations that are associated with those processes. To ensure that the dataset can be used in forecasting and now-casting, we prioritize data that is made available in near real-time or data that will be available in real-time on future missions. Figure 1 shows the complete set of inputs for the dataset, including any calculated parameters and the use of various event lists (see Section 3 for details on event lists). By considering all of the solar processes that can potentially alter SEP properties in interplanetary space (i.e., acceleration, transport, diffusion, among others), we create a whole-picture dataset that captures each observation's role in determining the occurrence probability and corresponding properties of the resulting SEP events.

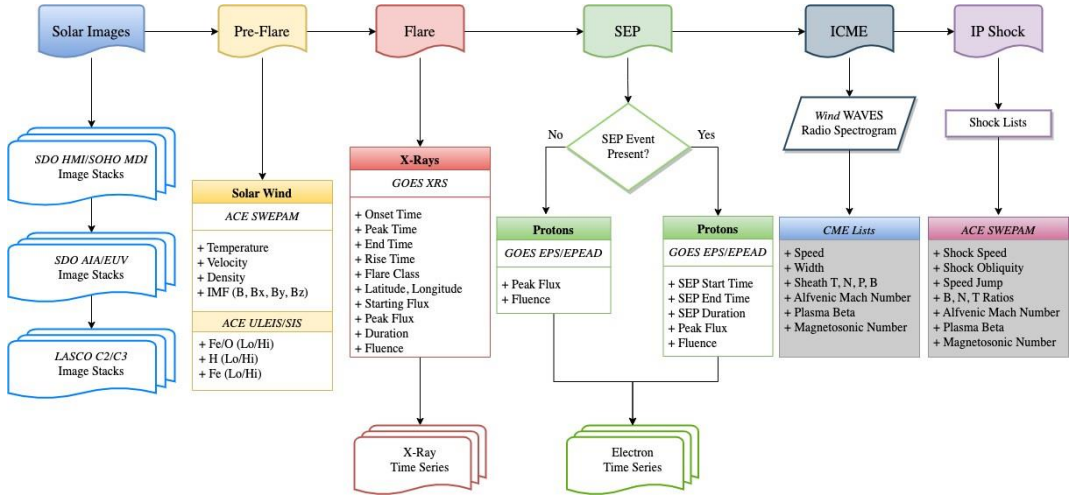

**Figure 1.** The complete dataset flowchart shows all incorporated observations from remote imaging to in-situ measurements. Each section relates to a time frame in the event series: Solar images (pre-flare, flare), solar wind conditions (pre-flare), X-ray properties, and time series (flare). Post-flare properties for the SEP event, Interplanetary Coronal Mass Ejection (ICME) parameters with Wind WAVES radio spectrogram, Interplanetary (IP) shock properties, and published shock lists. We note the instrumentation (in italics) used to obtain in-situ data along with each parameter observed or calculated parameter.

## 2 Instrumentation

Overall, we accumulated data from thirteen different instruments over five separate missions. Table 1 summarizes the spacecraft, instrument, measured observations, energy range, and time cadence.

### 2.1 Remote Sensing Observations

To capture the evolution and dynamics of active regions and to help inform ML models, we input remote observations of the solar atmosphere three days before the flare onset. We use the Michelson Doppler Imager (MDI; (Scherrer et al., 1995)) on board SoHO and the Helioseismic and Magnetic Imager (HMI; (Schou et al., 2012)) on board SDO to obtain full line-of-sight magnetograms to investigate the topology of the solar surface and solar atmosphere. Considering the possible memory limitation of GPUs, we downgrade the resolution of magnetograms to 256 pixels × 256 pixels and collect one magnetogram every 6 hours over 3 days before a flare onset. As a result of variable cadence and the instrument degradation to two images per day in the Extreme Ultraviolet Imaging Telescope (EIT;(Delaboudinière et al., 1995)) on board the Solar and Heliospheric Observatory (SoHO; (Domingo et al., 1995)), we use half-day (12 hour) resolution EUV data from 1996 onward. Starting in 2010, we used EUV data from the Atmospheric Imaing Assembly (AIA; (Lemen et al., 2012)) on board the Solar Dynamics Observatory (SDO; (Pesnell et al., 2012)) fixed at the same time cadence. To obtain radio observations, we make use of the WAVES instrument Radio Receiver Bands 1 and 2 (RAD1: 20 kHz - 1,040 kHz, RAD2: 1.075 MHz - 13.825 MHz) on board *Wind* ((Bougeret et al., 1995)) stationed at Lagrange Point 1 (L1). Using the twenty channels from RAD1 that contain observed data only (vs. interpolated) and a reduced set of sixty channels from RAD2 with observed data (eighty total frequency channels used to reduce data input size) at a down-sampled time of ten minutes using an anti-aliasing approach (low-pass filtering plus re-sampling), we create a single image of the logarithm of the intensities (see Figure 3). Finally, to observe the changes in the IP space as solar features such as CMEs streaming away from the sun, we obtain hourly images of the solar corona using the Large Angle and Spectrometric Coronagraph (LASCO; (Brueckner et al., 1995)) on board SoHO. Note that we do not fit the CME parameters such as width, propagation speed, duration, etc. In this work, we identify and connect the event parameters to known CMEs via published event lists. An image stack consists of various images corresponding to the instruments' cadence for each remote sensing source. See Section 2 and Figure 2 for more details. Data from these instruments is publicly available via JSOC web portals [1]. Solar imaging offers a large amount of information well-suited to ML algorithms. Keeping in mind data size, we implement data reduction techniques where needed and when appropriate while keeping any information loss to a minimum.

### 2.1.1 EUV

We choose the EUV band at 171 Å, which is a wavelength of a strong iron line (Fe IX) at approximately 600,000 Kelvin and observes changes in the corona and transition region boundary. This makes it ideal for studying the quiet corona and coronal loops, including fine plasma strands. To create a continuous dataset of EUV images (using SoHO and SDO) we download images of overlapping date and time, e.g. EIT 171 Å and AIA 171 Å at a cadence of 12 hours over 3 days before flare onset. We filter out the bad images using a 'QUALITY' keyword, normalize the images with exposure time, and correct for image degradation over time. We also perform degridding of the EIT/171 images. We then perform re-projection to homogenize images in terms of pixel scale, field of view, and point of view. The full-disc images are divided into patches, and the patches are aligned using a template-matching method. Finally, we use a low-resolution patch and limb darkening profile (encoding position of the patch pixels w.r.t. disc center) as input and the corresponding high-resolution patch as the output of a Deep Learning (DL) model presented in (Chatterjee et al., 2023). We use low-resolution versions (256 pixels × 256 pixels) of the homogenized images (DL model outcome) and interpolate for missing frames, creating the EUV data cube of size $256[pixels] \times 256[pixels] \times 7[frames]$ for each flare event. Figure 2a shows a sample EUV image and its corresponding stack size.

---

[1] http://jsoc.stanford.edu/

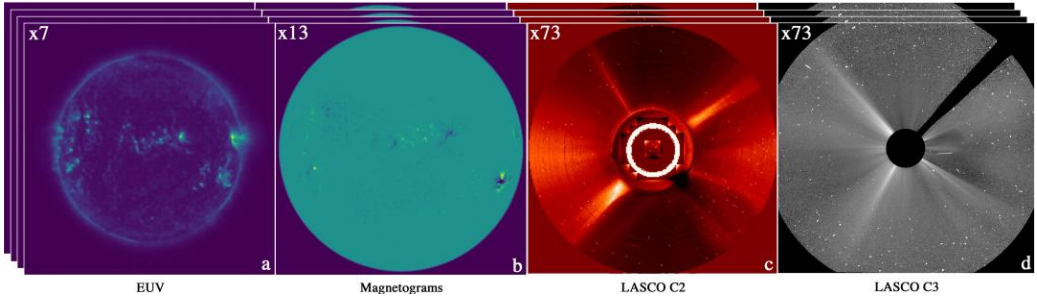

**Figure 2.** Imaging data is downloaded, and the sequence of images creates a stack that is ingested into the ML model. In this dataset, we use a time window beginning at the flare onset to 3 days prior. (a) The EUV stack contains 7 images (b) Magnetogram stack contains 13 images (c, d) LASCO C2 and C3 stacks contain 73 images each.

### 2.1.2 Magnetograms

Magnetograms capture the evolution of sunspots and spatial variations in the solar magnetic field properties, especially the surface distribution and polarity of those magnetic fields. SoHO/MDI and SDO/HMI full-disc line-of-sight magnetograms are another imaging input feature of our dataset, and a sample event is shown in Figure 2b. Keeping in mind the memory limitation of most graphics processing units (GPUs) we downgrade the resolution of magnetograms to 256 pixels × 256 pixels and collect one magnetogram every 6 hours over 3 days before the flare onset. This creates a data cube of size $256[pixels] \times 256[pixels] \times 13[frames]$. Note the original size is 1024x1024 for MDI and 4096x4096 for HMI, with a cadence of 96 minutes and 45 seconds, respectively. This image downsampling reduces the computer's memory burden without losing information about the active region evolution. Considering Convolutional Neural Networks (CNNs) are blind to the time stamps of input magnetograms, we ensure every two consecutive magnetograms are separated by 6 hours by performing nearest-neighbor interpolation for each pixel over the time axis. To homogenize MDI and HMI magnetograms, we use a conversion factor of 1.3 (MDI = 1.3*HMI) and clip the field strength within [-1000 Gauss, 1000 Gauss]. This creates a continuous dataset for longer observation periods than a single instrument would provide. Finally, we normalize the field strength ($F$) using the transformation $\frac{1}{2}(1 + \frac{f}{1000})$. So, a pixel value of 0, 0.5, and 1 in the normalized magnetograms represent -1000 Gauss, 0 Gauss, and 1000 Gauss.

### 2.1.3 LASCO C2 and C3

Coronagraphs block the solar disk using an external occulter, revealing coronal features such as coronal streamers and CMEs and their propagation into the interplanetary medium. SOHO has two onboard coronagraphs, C2, which has a 3-degree field of view (1.5 to 6 solar radii), and C3, which has a 16-degree field of view (3.7 to 30 solar radii). Sample event images for the LASCO C2 and C3 images can be seen in Figures 2c and 2d, respectively. We download level 0.5 LASCO - C2 and C3 images using JSOC at a cadence of 1 hour over a period of 3 days before flare onset. We apply the SSWIDL (SolarSoft) package to convert level 0.5 to level 1 and apply a normalizing radial graded filter to equalize the contrast of coronal structures with respect to the background at different radial distances. Finally, we down-sample those images to a fixed size of 256 pixels × 256 pixels and perform a nearest neighbor interpolation for each pixel of down-sampled images over the time axis to fill missing frames. This generates a C2 and C3 data cube of size $256[pixels] \times 256[pixels] \times 73[frames]$ for each flare event.

### 2.1.4 Wind WAVES

Radio burst signatures have been shown to correlate with SEP events (Gopalswamy et al., 2008; Cane et al., 2002). Type II radio bursts can indicate particle acceleration from CME-driven shocks, while Type III radio bursts are associated with accelerated electrons leaving the sun along open magnetic field lines. The *Wind* Radio and Plasma Wave Experiment (WAVES) instrument provides comprehensive coverage of radio and plasma wave phenomena in the frequency range of 20 kHz up to 13.825 MHz. We can identify radio burst signatures of accelerated particles using Radio Receiver Bands 1 and 2 (RAD1: 20 kHz - 1,040 kHz; RAD2: 1.075 MHz - 13.825 MHz). Only 20 channels of RAD1 contain observed data (the remaining channels are interpolated), and all 256 channels of RAD2 contain observed data; however, to reduce the data input, we use a total of 80 frequency channels from RAD1 and RAD2. We down-sample the time axis to a cadence of 10 minutes, take the logarithm of the intensities (to prevent CNNs from being driven by outlier pixels), and clip them within the range [-1,1], creating a single image of WAVES data 3-days before the flare onset in the form of a 432 time-frequency radio image. Figure 3 shows a 2D histogram of frequency observations from *Wind* between Jan. 20, 2005, and Jan. 23, 2005. At the beginning of the time interval, there is a clear type III radio burst.

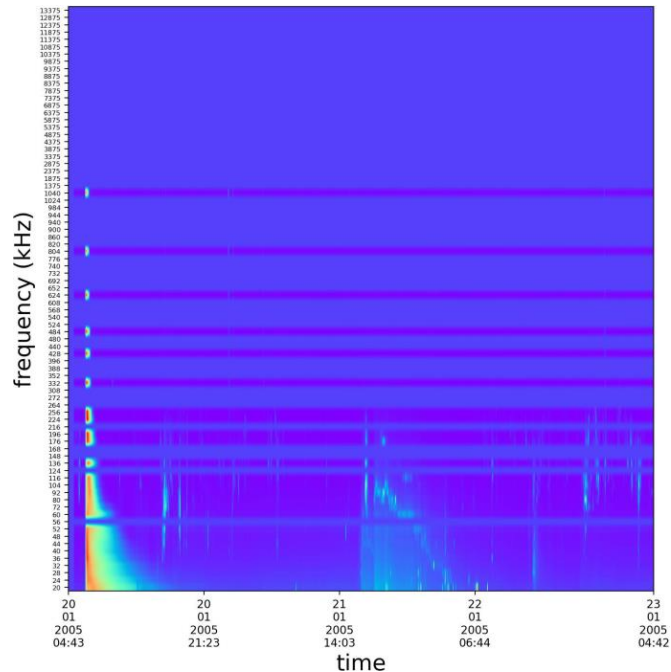

**Figure 3.** *Wind* WAVES sample event image showing a Type III solar radio burst, evidence of flare accelerated electrons.

## 2.2 In-situ Observations

In addition to remote observations, we provide a wide variety of near-Earth *in-situ* plasma measurements before, during, and after each observed flare event. We complement the GOES Flare

Event List by measuring the integrated high-energy protons ($\geq$ 5 MeV, $\geq$ 10 MeV, $\geq$ 30 MeV, $\geq$ 60 MeV, $\geq$ 100 MeV) using the GOES Energetic Particle Sensor (EPS; (Sellers & Hanser, 1996)), after 2017 the Solar and Galactic Proton

Sensor (SGPS) on the Space Environment In Situ Suite (SEISS). In addition to high- energy protons, we include the proton, alpha, and heavy ion fluxes in the suprathermal energy range (0.05 - 5.0 MeV). We measure the low-energy proton fluxes every 12 seconds with the Energetic Proton and Alpha Monitor (EPAM; (Gold et al., 1998)), and we measure the hourly-averaged alpha and heavy ion fluxes using the Ultra-Low Energy

Ion Spectrometer (ULEIS; (Mason et al., 1998)) and the Solar Isotope Spectrometer (SIS; (Stone et al., 1998)). These instruments are part of a suite on board the Advanced Composition Explorer (ACE; (Stone et al., 1998)), also stationed at L1. We also obtain the electron time series data at energies of 0.038 - .315 MeV from EPAM. For solar wind properties (i.e. velocity, temperature, density, magnetic field configuration), we utilize two instruments on board ACE; the local particle population data is from the Solar Wind Electron, Proton, and Alpha Monitor (SWEPAM; (McComas et al., 1998)) with a 64- second time cadence, and local magnetic field vector properties from the Magnetic Field Experiment (MAG; (Smith et al., 1998)) with a 16-second time cadence.

| Mission | Instrument | Measurement | Energy Range | Resolution |
|---------|-----------|-------------|--------------|------------|
| SOHO | LASCO | Solar Corona (C2, C3) | | 1 hour |
| | MDI | Line of Sight Solar Magnetic Field | | 6 hour |
| | EIT | Solar Extreme UV Emissions | 171 Å | 12 hour |
| SDO | HMI | Line of Sight Solar Magnetic Field | | 6 hour |
| | AIA | Solar Extreme UV Emissions | 171 Å | 12 hour |
| *Wind* | WAVES | Solar Radio Emissions | 10 kHz - 10 MHz | 0.1 second |
| GOES | EPS | $H^+$ and $e^-$ | 0.6 - 500 MeV | 5 minute |
| | XRS | X-Ray | 0.5 - 8.0 Å | 3 second |
| ACE | EPAM | $H^+$ | 0.05 - 5 MeV | 12 second |
| | ULEIS | $^3$He, $^4$He, O, Fe | 0.04 - 10 MeV | 1 hour |
| | SIS | $^3$He, $^4$He, O, Fe | 3 - 170 MeV | 1 hour |
| | SWEPAM | Solar Wind Protons | 0.4 - 4 keV | 1 minute |
| | MAG | Local Magnetic Field | | 16 second |

**Table 1.** Complete instrumentation table.

### 2.2.1 Solar Wind

Solar wind conditions arrive at Earth sometimes delayed by several days after energetic solar events; however, in-situ measurements are critical to gauging the conditions of the solar atmosphere and interplanetary medium associated with these events. We include the first three solar wind moments (i.e., density, velocity, and temperature) and the vector components and magnitude of the Interplanetary Magnetic Field (IMF). Turbulence in the interplanetary medium is reflected in the corresponding 1 au solar wind measurements, which may alter particle acceleration and transport of said events. Thus, the 24-hour averaged solar wind conditions provide insight into pre-flare solar activity, solar cycle phase, and expected route to 1 au. Figure 4, panels e & f show the solar wind speed, density, and temperature. The yellow highlighted region shows the time window used for parameter calculations. Panel g shows the magnetic field magnitude and vector quantities. The pre-flare conditions are highlighted in purple, and the ICME signature is highlighted in grey.

### 2.2.2 Suprathermal Population

Suprathermal particles are an extension of the solar wind tail at higher energies. These particles have energies ranging from a few keV to a few MeV. The origin of this population of particles is presently debatable (Tylka et al., 2005; Mason & Sanderson, 1999; Cane et al., 2006). They could be scattered remnants of large-scale transient events (Dayeh et al., 2017), or they could be accelerated stochastically in the solar atmosphere or in localized regions of high density in the solar wind (Fisk & Gloeckler, 2006). Nevertheless, they are nearly permanently present in the inner heliosphere, and they play an important role in post-flare SEP enhancement.

It is theorized that the highest energy particles in any given event already have large initial energies before the event. They are then exponentially re-accelerated by turbulence at the flare site or by propagating shocks (Fermi, 1949; Reames, 2017; M. I. Desai et al., 2007; Giacalone, 2005). Because of this phenomenon, we believe the $\geq$ 10 MeV time profiles strongly depend on the suprathermal seed population just before the flare onset (Kahler, 2001; E. Cliver, 2006). Therefore, we include the average pre-flare fluxes and abundance ratios of three of the most abundant energetic particle species (H, O, & Fe), which, to date, are not typically used in ML models. Panel d, Figure 4 shows the Oxygen (O) particle flux during a sample event, and the highlighted (cyan) area shows the sampling window used to calculate the parameters (i.e. Fe/O ratio) that are saved to the dataset. The Hydrogen (H), Iron (Fe) abundances and ratios are done using the same sampling window.

### 2.2.3 X-Rays

Magnetic reconnection converts magnetic energy into particle kinetic energy. Particles accelerated down closed loops collide with the chromosphere and decelerate, producing large amounts of X-rays via Bremsstrahlung collisions (Galloway, R. K. et al., 2010).
X-rays can cause issues such as radio blackout or increased satellite drag (Yasyukevich et al., 2018; Xiong et al., 2014). X-rays are observed in the 0.05 - 0.4nm and 0.1 - 0.8nm wavelength bands by a primary and a secondary GOES satellite at a cadence of 60 seconds. We use the 0.1 - 0.8nm wavelength band (also known as XL, the black line in Figure 4, panel a) to calculate the starting flux, peak flux (marked by the red star in Figure 4, panel a), rise time, fluence, and duration for each flare event. X-rays have been used in many SEP models including the Space Weather Prediction Center (SWPC), Air Force Research Laboratory (AFRL), etc. ((Balch, 2008; Kahler & Ling, 2015),). Flare class prediction has been used in Support Vector Machines (SVMs) (Choi et al., 2012; Nishizuka et al., 2018; Bobra & Couvidat, 2015) and was used as a predictive output of one of the first attempts in neural networks by Fozzard (1989). There has been success using Long Short-Term Memory (LSTM) to predict X-ray time profiles (Li et al., 2020), and we find adding the X-ray time series beneficial to the ML model (orange highlighted area of Figure 4, panel a).

### 2.2.4 High-Energy Protons & Electrons

Isotropic and permanently present high-energy protons in the heliosphere at energies above 10 MeV typically originate beyond the heliosphere as low-energy cosmic rays and their fluxes are too low to be considered a serious threat. However, strong X-ray flares on the Sun can produce CMEs and shocks that efficiently accelerate protons to high energies and significantly increase radiation levels (Shea & Smart, 2012; Reames, 2013). This type of particulate radiation is of paramount concern and is consequently our primary measurement in determining event observations (see Section **4**: SEP Event Detection). Using GOES EPS with a 5-minute time cadence, we determine the onset time of the proton enhancement begins when three consecutive time steps are above the pre-determined threshold and its peak time (Figure 4 panel b, red x), end time, and total fluence. These values are included in the dataset. Figure 4, panel b highlights (green) the time window for calculating the SEP parameters for a SEP-positive event. High energy electrons, though not energetic enough to pose a serious threat to human health can cause space craft charging and damage electronics. These MeV electrons arrive before the protons and are a strong predictor of the subsequent proton enhancement (Posner, 2007). We, therefore, include the electron time series measurements taken at L1 on ACE EPAM 24 hours before the flare onset.

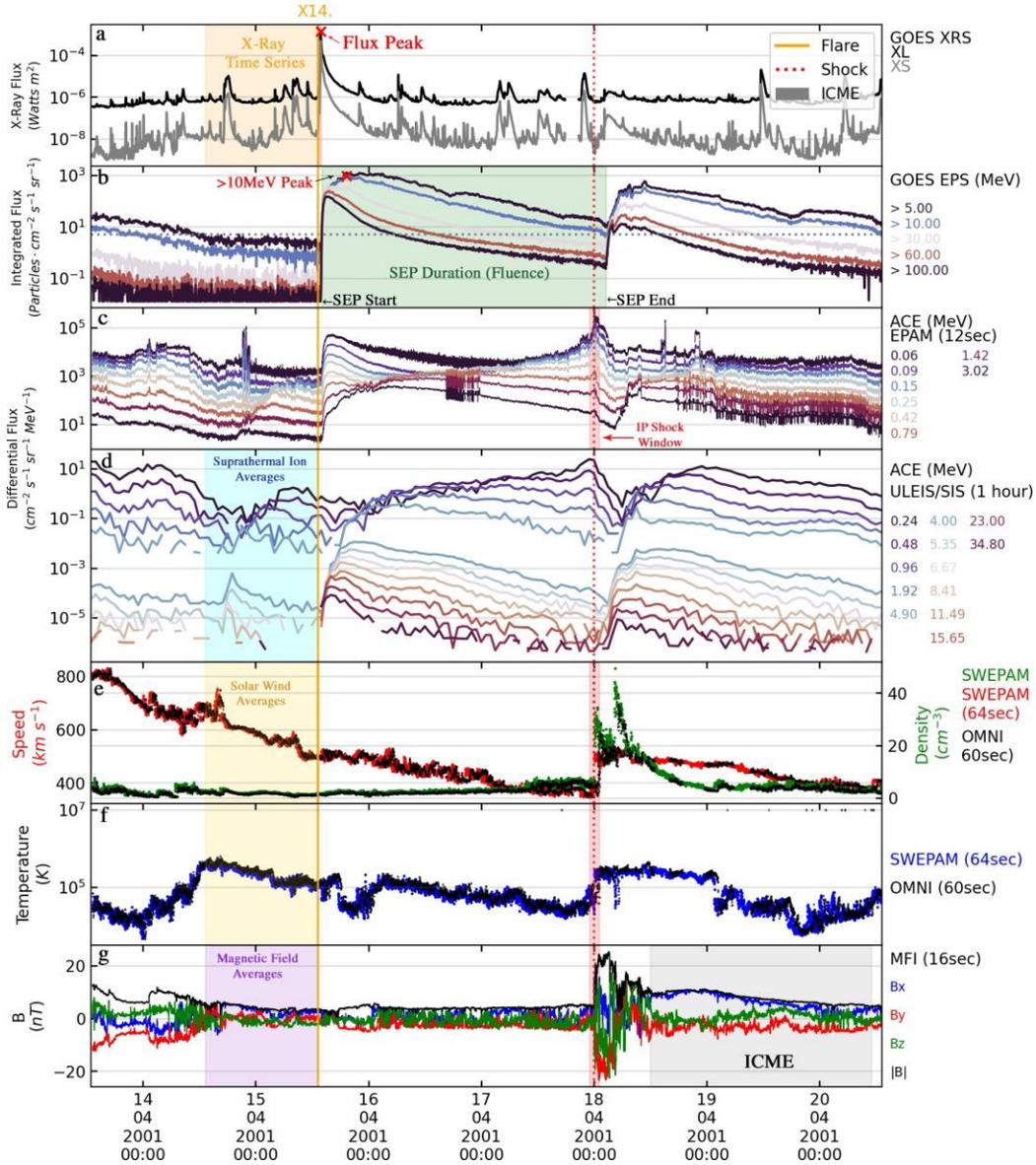

**Figure 4.** This plot of a sample SEP-positive event in the dataset. Each panel pertains to a different instrument or observation. Panel (a) contains the X-ray flux, (b) shows the corrected integrated flux from GOES used to determine if a SEP event occurs and its subsequent properties,

(c) at L1, we use proton flux from ACE to calculate properties during the shock, (d) including H, O, and Fe in low and high energies and ratios allows for observations of the suprathermal ion population before the event, (e,f) the pre-flare solar wind data is calculated in the pre-flare time window show in the yellow highlight, (g) the magnitude and vectors of the interplanetary magnetic field are averaged and ICME properties are shown.

## 3  Event Lists

The National Oceanic and Atmospheric Administration (NOAA) began operating the GOES satellites with onboard X-ray and energetic particle detectors in 1975. Since then, all detected solar flares have been cataloged by NOAA and are publicly available through the Heliophysics Events Knowledgebase (HEK). This list contains the GOES flare classification (e.g., X, M, C, B, A) and the flare properties such as start time, peak, and end time. When available, the list provides the flare location (latitude, longitude) and the corresponding active region (AR) number; flare events that happen beyond the solar limbs may not contain location or AR information. We obtain this list using SunPy's pre-built query of the GOES flare event list through the HEK (Barnes et al., 2020). Figure 5 shows a sample time series of the long (1.0 - 8.0 Å) and short (0.5 - 3.0 Å) X-ray bands observed by GOES with flares identified from the list. An explanation of the GOES event list flare detection algorithm is detailed in Ryan et al. (2016).

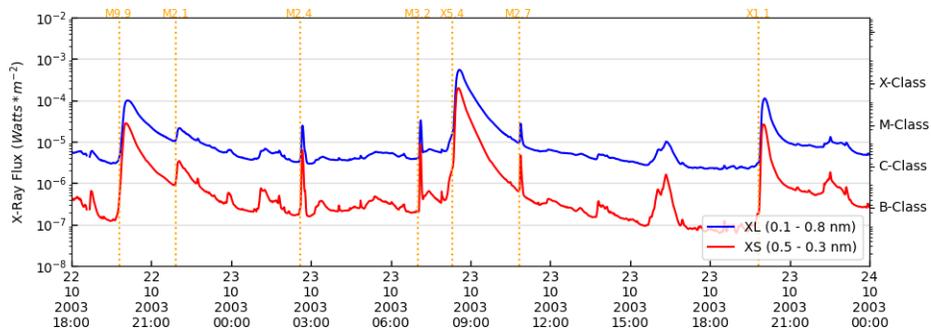

**Figure 5.**  Sample X-ray time series from GOES flare event list with flare onset time marked by the dashed orange vertical line and flare class identifier.

Interplanetary shock data is integrated from the Database of Heliospheric Shockwaves maintained at the University of Helsinki (Kilpua et al., 2015), which identifies over 2,600 shocks detected at ACE, WIND, and STEREO from 1975 to the present. The red vertical dashed line in Figure 4 marks an IP shock arrival during a sample event from the dataset. CME parameters compiled since 1996 by Cane and Richardson (2003) are also ingested into the data set, and an example CME event is highlighted in gray in Figure 4, panel g. For events after 2010, we can incorporate the Community Coordinated Modeling Center (CCMC) Space Weather Database of Notifications, Knowledge, Information (DONKI) web service to determine and verify linked events.

## 4  SEP Event Detection

We start the SEP event identification process by identifying a single flare in the NOAA/GOES flare event list discussed in section **3**. We then look for an enhancement in the $\geq 10$ MeV corrected integrated flux within 6 hours of the flare onset. If the flux of the proton enhancement exceeds 5 pfu (1 pfu = 1 proton cm$^{-2}$ s$^{-1}$ sr$^{-1}$) for 15 minutes (3 consecutive data points) and the maximum pre-flare background flux (maximum flux value in the $\geq 10$ MeV over a 3-hour time window before flare onset shown in the red highlight of Figure 5), then the event is considered a SEP-positive event. Figure 6 shows the profile of the flare and the SEP event for an event marked as SEP-positive.
If the proton flux does not rise above the 5 pfu threshold, it is considered a negative event. In the event of multiple flares in a short time frame, the proton enhancement is attributed to the flare with the largest peak X-ray flux. Any events where the peak X-ray flux data is unavailable are removed from the dataset. We choose a conservative value of 5 pfu as the threshold energy instead of the typical 10 pfu, given by the NOAA SWPC definition for a proton event, because we are aiming for a conservative flux value that optimizes the number of positive events while still lying well above the instrument's background signal.

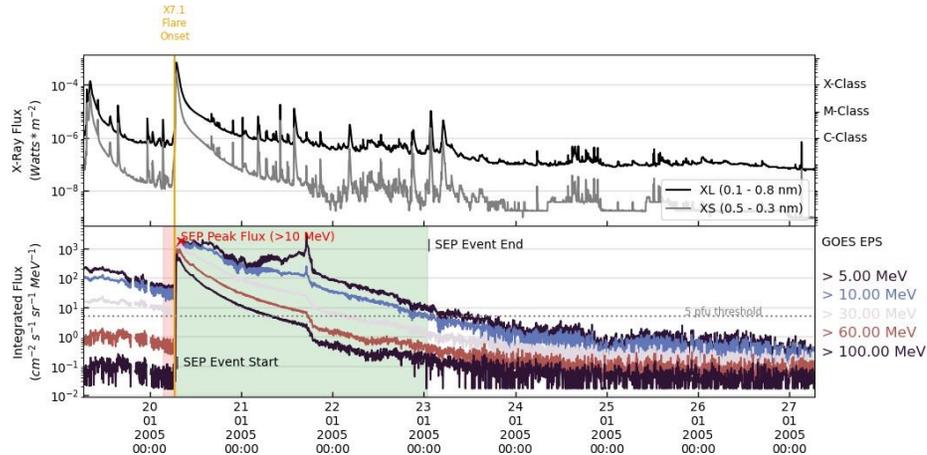

**Figure 6.** GOES X-ray and proton time series during a flare event marked as SEP-positive. The vertical orange line marks the X7.1 flare onset, and the green box highlights the associated SEP event duration, the time frame the fluence is calculated over. The red X marks the peak flux value in the $\geq 10$ MeV energy, and the 5 pfu threshold is marked by the horizontal dotted grey line. The red highlighted area is the time to calculate the pre-flare flux values. To be positive,
the SEP enhancement must be larger than the max flux value pre-flare.

For positive SEP events, the start and stop time of the SEP are determined by the time the $\geq 10$ MeV flux rises about the threshold for three consecutive points (15 minutes) and then falls below the threshold for three consecutive points (15 minutes). We then use the SEP start and stop time to calculate the duration of the SEP event and the time from flare onset to SEP onset. For SEP-negative events (Figure 7) these parameters are filled with -9999. The peak flux and fluence at five different energy ranges ($\geq 5$ MeV, $\geq 10$ MeV, $\geq 30$ MeV, $\geq 60$ MeV, $\geq 100$ MeV) are calculated for SEP-positive events over the duration of the event; for SEP-negative events, these values are calculated over a pre-determined time window of 6 hours (shown in the green highlight area of Figure 7.

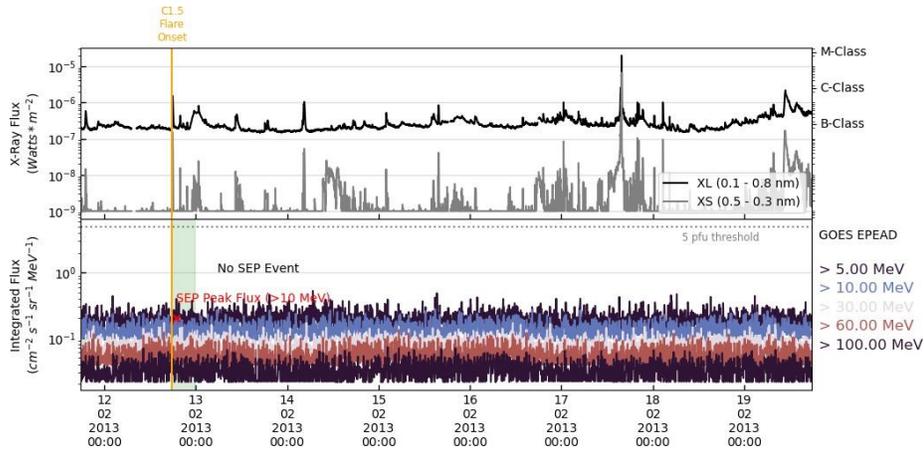

**Figure 7.** GOES X-ray and proton time series during a flare event marked as SEP-negative. The vertical orange line marks the C1.5 flare onset. The ≥ 10 MeV flux does not exceed the 5 pfu threshold marked by the dotted grey line during the preset time frame. Included values are calculated in the 6-hour window marked by the green highlighted area.

We repeat this process for each flare in the list between 1998 - 2013 and categorize them by parent flare class. Of the 131 X-class flares, 41 are positive (31.3%). 71 out of 1649 M-class flares are positive (4.3%), and 140 out of 16014 C-class flares are positive (0.87%). Note the positive percentage rate increases with increasing flare class, as shown in Figure 8. This trend is consistent with the known correlations between flare strength and SEP enhancement (Kahler & Ling, 2018).

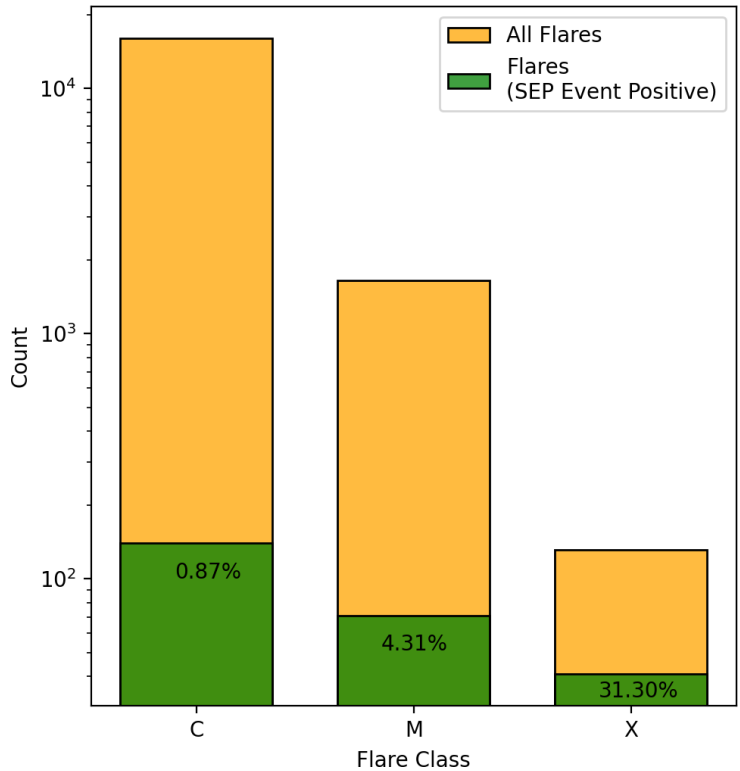

**Figure 8.** Class distribution of all GOES solar flare events (orange) by flare class from 1998- 2013. The subset of SEP-positive flare events is shown in green, and their percentage is noted in the respective bar. While the total number of SEP-positive flare events decreases with increasing flare class, the percentage of SEP-positive flare events with respect to the total number of flares in each class increases.

The association between sunspot number and SEP events has been well studied (i.e., (Birch & Bromage, 2022), (Marroquin et al., 2023), (Barnard & Lockwood, 2011)). Our dataset shows a similar correlation between the number of SEP-positive events and the sunspot numbers reported by NOAA SWPC during SC 23 and SC 24. When considering larger SEP events ( $\geq$ 10 MeV particle flux exceeds the 10 pfu threshold), the pattern continues as shown in Figure 9. These findings reinforce the influence of solar activity, indicated by sunspot numbers, on the occurrence and magnitude of SEP events.

Flares originating from active regions west of the solar disk's central meridian are more likely to produce impactful SEPs near the Earth owing to Earth's magnetic connectivity to the CME-driven shock in or just beyond the solar corona. Figure 10 illustrates the flare location on the solar disk compared with the logarithmic intensity of the $\geq$ 10 MeV particle energies using our dataset.

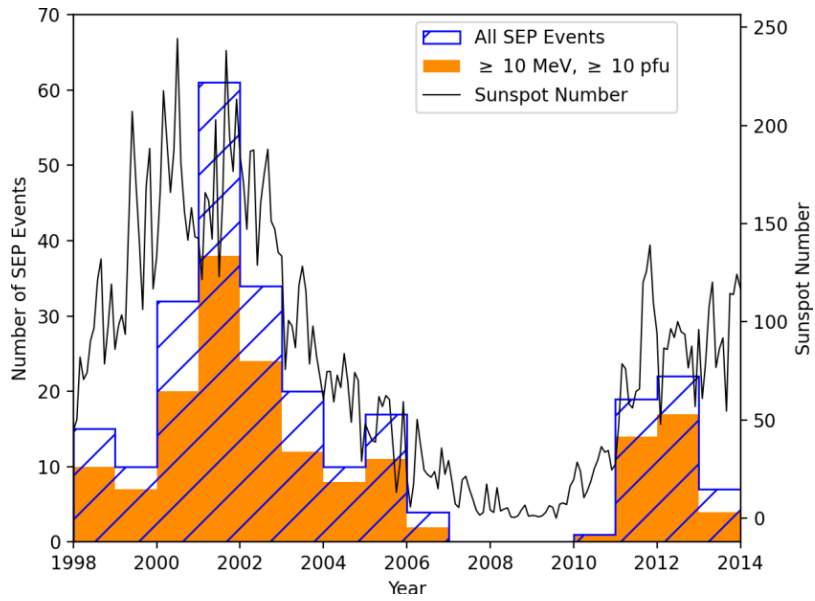

**Figure 9.** Histogram of the annual number of all SEP-Positive events (blue hashed bins) and SEP-positive events where the ≥ 10 MeV protons cross the 10 pfu threshold (filled orange bins). Monthly sunspot numbers are over-plotted in black. We find that the number of SEP-positive events and the subset of strong SEP-positive events follow the sunspot cycle.

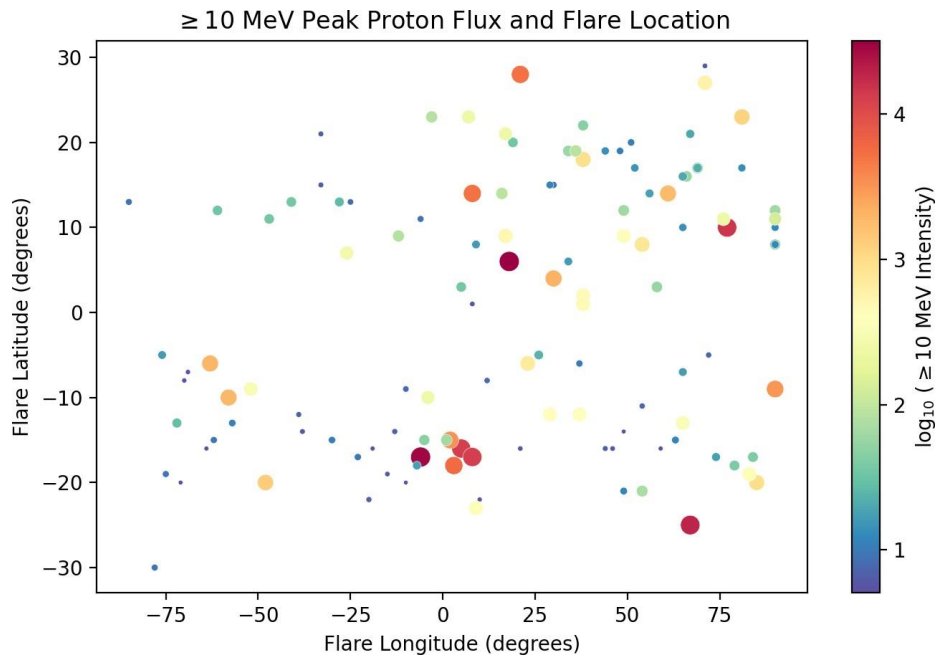

**Figure 10.** The solar disk location of each SEP-positive flare. The color and size indicate the logarithm of the ≥ 10 MeV proton intensity, indicated by the color bar. SEP-positive events tend to occur when the flare

originates west of the central meridian, and the average intensity of the SEP-positive events is greater in the western hemisphere compared with the eastern hemisphere.

## 5 ML/AI-Ready Data

To produce accurate and reliable model results, it is important to provide researchers using ML models with a carefully validated dataset. We do this by considering the caveats of each instrument measurement and identifying and correcting any errors or non-valid observations, and ensuring that the data is not noisy, inaccurate, or biased. The first step we take in the data process is verifying primary and secondary GOES satellites. The dates for each satellite's primary or secondary designation can be found on the NOAA documentation website [2]. Second, we identify and fill any non-valid data for each instrument and set these values to -9999, or in cases where the missing data is used for calculating a vital parameter, the event is removed from the dataset. Lastly, we provide consistent sampling by interpolating data when it is appropriate to do so. The data is labeled according to each process it belongs to (flare, pre-flare, SEP, etc.), saved, and made available in formats easily readable by ML algorithms.

## 6 Future Work

Events from Solar Cycle 24 and 25, spanning the years 2013-2020, are being added to the next iteration of the dataset. Parameters from the lists described in section 3 are carefully linked to their parent event. The addition of meaningful parameters will continue, and their data description will be added as the dataset matures.

## 7 Open Research

*Software:* This dataset used version 4.0 of SunPy open source software package (Barnes et al., 2020), version 3.5.1 matplotlib (Hunter, 2007), version 1.21.2 NumPy (Harris et al., 2020), version 1.4.1 pandas (pandas development team, 2020), version 1.7.2 SciPy (Virtanen et al., 2020), version 0.11.2 seaborn (Waskom, 2021), and customized download scripts written by Samuel Hart.

*Data Availability Statement:* The 'X-ray Flare' dataset and the GOES X-ray sensor data were prepared and made available through the NOAA National Geophysical Data Center (NGDC) [3]. We acknowledge the use of NOAA's 1–8 Å solar X-ray data and proton data from the NOAA GOES data archive. The MDI and AIA data is used courtesy of NASA/SDO and the AIA, EVE, and HMI science teams [4]. The SOHO/LASCO data used here are produced by a consortium of the Naval Research Laboratory (USA) Max- Planck-Institut fuer Aeronomie (Germany), Laboratoire d'Astronomie (France), and the University of Birmingham (UK)[5]. SOHO is a project of international cooperation between ESA and NASA. This paper uses data from the Heliospheric Shock Database, generated and maintained at the University of Helsinki[6].

This dataset is available on Zenodo under a Creative Commons Attribution license at doi: 10.5281/zenodo.10044864. Customization scripts will also be added to allow users to set variables and create their own dataset.

---

[2] https://ngdc.noaa.gov/stp/satellite/goes/documentation.html
[3] https://www.ngdc.noaa.gov/stp/satellite/goes/
[4] https://sdo.gsfc.nasa.gov/data/aiahmi/
[5]     https://www.cosmos.esa.int/web/soho/soho-science-archive
[6] http://ipshocks.fi/

**Acknowledgments**

This project has been mainly supported by an Operations 2 Research grant number 80NSSC20K0290. Partial support was also provided by O2R 80HQTR20C0017, Heliophysics Living With a Star Science Program, under grant number 80NSSC19K0079.

# References

Angryk, R. A., Martens, P. C., Aydin, B., Kempton, D., Mahajan, S. S., Basodi, S., . . . Georgoulis, M. K. (2020). Multivariate time series dataset for space weather data analytics. *Scientific Data*, *7* (1), 227. Retrieved from https:// doi.org/10.1038/s41597-020-0548-x doi: 10.1038/s41597-020-0548-x

Balch, C. C. (2008). Updated verification of the space weather prediction center's solar energetic particle prediction model. *Space Weather*, *6* (1). Retrieved from https://agupubs.onlinelibrary.wiley.com/doi/abs/10.1029/ 2007SW000337 doi: https://doi.org/10.1029/2007SW000337

Barnard, L., & Lockwood, M. (2011). A survey of gradual solar energetic particle events. *Journal of Geophysical Research: Space Physics*, *116* (A5). Retrieved from https://agupubs.onlinelibrary.wiley.com/doi/abs/10.1029/ 2010JA016133 doi: https://doi.org/10.1029/2010JA016133

Barnes, W. T., Bobra, M. G., Christe, S. D., Freij, N., Hayes, L. A., Ireland, J., . . . Dang, T. K. (2020). *The Astrophysical Journal*, *890*, 68-. Retrieved from https://iopscience .iop.org/article/10.3847/1538-4357/ab4f7a doi: 10.3847/1538-4357/ ab4f7a

Birch, M., & Bromage, B. (2022). Sunspot numbers and proton events in solar cycles 19 to 24. *Journal of Atmospheric and Solar-Terrestrial Physics*, *236*, 105891. Retrieved from https://www.sciencedirect.com/science/article/ pii/S1364682622000657 doi: https://doi.org/10.1016/j.jastp.2022.105891

Bobra, M. G., & Couvidat, S. (2015, jan). Solar flare prediction using sdo/hmi vector magnetic field data with a machine-learning algorithm. *The Astrophys- ical Journal*, *798* (2), 135. Retrieved from https://doi.org/10.1088%2F0004 -637x%2F798%2F2%2F135 doi: 10.1088/0004-637x/798/2/135

Bobra, M. G., Wright, P. J., Sun, X., & Turmon, M. J. (2021). SMARPs and SHARPs: Two Solar Cycles of Active Region Data. *The Astrophysical Journal Supplement Series*, *256* (2), 26. doi: 10.3847/1538-4365/ac1f1d

Bougeret, J. L., Kaiser, M. L., Kellogg, P. J., Manning, R., Goetz, K., Monson, S. J., . . . Hoang, S. (1995). WAVES: The radio and plasma wave investiga- tion on the wind spacecraft. *Space Science Reviews*, *71* (1-4), 231–263. doi: 10.1007/BF00751331

Brueckner, G., Howard, R., Koomen, M., Korendyke, C., Michels, D., Moses, J., . . . Eyles, C. (1995). Visible Light Coronal Imaging And Spectroscopy. *Solar Physics*, *162* (1-2), 357–402.

Cairns, I. H., Lobzin, V. V., Donea, A., Tingay, S. J., McCauley, P. I., Oberoi, D., . . . Williams, C. L. (2018). Low altitude solar magnetic reconnection, type iii solar radio bursts, and x-ray emissions. *Scientific reports*(1), 1612–1676. doi: 10.1038/s41598-018-19195-3

Camporeale, E. (2019). The Challenge of Machine Learning in Space Weather: Now- casting and Forecasting. *Space Weather*, *17* (8), 1166–1207. doi: 10.1029/ 2018SW002061

Cane, H. V., Erickson, W. C., & Prestage, N. P. (2002). Solar flares, type iii radio bursts, coronal mass ejections, and energetic particles. *Journal of Geophys- ical Research: Space Physics*, *107* (A10), SSH 14-1-SSH 14-19. Retrieved from https://agupubs.onlinelibrary.wiley.com/doi/abs/10.1029/ 2001JA000320 doi: https://doi.org/10.1029/2001JA000320

Cane, H. V., Mewaldt, R. A., Cohen, C. M. S., & von Rosenvinge, T. T. (2006). Role of flares and shocks in determining solar energetic particle abundances. *Journal of Geophysical Research: Space Physics*, *111* (A6). Retrieved from https://agupubs.onlinelibrary.wiley.com/doi/abs/10.1029/ 2005JA011071 doi: https://doi.org/10.1029/2005JA011071

Cane, H. V., & Richardson, I. G. (2003). Interplanetary coronal mass ejec- tions in the near-earth solar wind during 1996–2002. *Journal of Geo- physical Research: Space Physics*, *108* (A4). Retrieved from https:// agupubs.onlinelibrary.wiley.com/doi/abs/10.1029/2002JA009817 doi: https://doi.org/10.1029/2002JA009817

Chancellor, J. C., Scott, G. B., & Sutton, J. P. (2014). Space radiation: The number one risk to astronaut health beyond low earth orbit. *Life*, *4* (3), 491–510. doi: 10.3390/life4030491

Chatterjee, S., Dayeh, M., Muñoz-Jaramillo, A., Bain, H. M., Moreland, K., & Hart, S. (2023). Mempsep i : Forecasting the probability of solar energetic par- ticle event occurrence using a multivariate ensemble of convolutional neural networks. *Space Weather* .

Chatterjee, S., Muñoz-Jaramillo, A., Dayeh, M. A., Bain, H. M., & Moreland, K. (2023, September). Homogenizing SOHO/EIT and SDO/AIA 171 Å Images: A Deep-learning Approach. , *268* (1), 33. doi: 10.3847/1538-4365/ace9d7

Choi, J.-H., Choi, C., Ko, B.-K., & Kim, P.-K. (2012). Detection of Cross Site Scripting Attack in Wireless Networks Using n-Gram and SVM. *Mobile infor- mation systems*, *8* (3), 275–286. doi: 10.3233/MIS-2012-0143

Cliver, E. (2006). The unusual relativistic solar proton events of 1979 august 21 and 1981 may 10. *The Astrophysical Journal* , *639* (2), 1206.

Cliver, E. W., Dennis, B. R., Kiplinger, A. L., Kane, S. R., Neidig, D. F., Sheeley, J., N R, & Koomen, M. J. (1986). Solar gradual hard X-ray bursts and associ- ated phenomena. *The Astrophysical Journal* , *305* , 920. doi: 10.1086/164306

Crosby, N., Heynderickx, D., Jiggens, P., Aran, A., Sanahuja, B., Truscott, P., . . . Hilgers, A. (2015). Sepem: A tool for statistical modeling the solar en- ergetic particle environment. *Space Weather* , *13* (7), 406-426. Retrieved from https://agupubs.onlinelibrary.wiley.com/doi/abs/10.1002/ 2013SW001008 doi: https://doi.org/10.1002/2013SW001008

Dayeh, M. A., Chatterjee, S., Munoz-Jaramillo, A., Moreland, K., Bain, H. M., & Hart, S. (2023). Mempsep ii. – forecasting the properties of solar energetic particle events using a multivariate ensemble approach. *Space Weather* .

Dayeh, M. A., Desai, M. I., Mason, G. M., Ebert, R. W., & Farahat, A. (2017). Origin and properties of quiet-time 0.11–1.28 mev nucleon1 heavy-ion population near 1 au. *The Astrophysical Journal* , *835* (2), 155. doi:10.3847/1538-4357/835/2/155

Decker, R. B. (1981). The modulation of low-energy proton distributions by propagating interplanetary shock waves: A numerical simulation. *Jour- nal of Geophysical Research: Space Physics*, *86* (A6), 4537–4554. doi: 10.1029/JA086iA06p04537

Delaboudinière, J. P., Artzner, G. E., Brunaud, J., Gabriel, A. H., Hochedez, J. F., Millier, F., . . . Van Dessel, E. L. (1995). EIT: Extreme-ultraviolet Imaging Telescope for the SOHO mission. *Solar Physics*, *162* (1-2), 291–312. doi: 10.1007/BF00733432

Desai, M., & Giacalone, J. (2016). Large gradual solar energetic particle events. *Liv- ing Reviews in Solar Physics*, *13* (1). doi: 10.1007/s41116-016-0002-5

Desai, M. I., Mason, G. M., Gold, R. E., Krimigis, S. M., Cohen, C. M. S., Mewaldt, R. A., . . . Dwyer, J. R. (2007). Evidence for a Two-Stage Acceleration Process in Large Solar Energetic Particle Events. *Space Science Reviews*, *130* (1), 243– 253. Retrieved from https://doi.org/10.1007/s11214-007-9219-x doi: 10.1007/s11214-

007-9219-x

Domingo, V., Fleck, B., & Poland, A. I. (1995). SOHO: The Solar and He- liospheric Observatory. *Space Science Reviews*, *72* (1-2), 81–84. doi: 10.1007/BF00768758

Fermi, E. (1949). On the Origin of the Cosmic Radiation. *Physical review*, *75* (8), 1169–1174. doi: 10.1103/PhysRev.75.1169

Fisk, L. A., & Gloeckler, G. (2006, feb). The common spectrum for accelerated ions in the quiet-time solar wind. *The Astrophysical Journal*, *640* (1), L79. Re- trieved from https://dx.doi.org/10.1086/503293 doi: 10.1086/503293

Fozzard, R. (1989). *Theonet: a connectionist expert system for solar flare forecasting* (Doctoral dissertation, University of Colorado, Boulder, Department of Com- puter Science). Retrieved from https://spl.cde.state.co.us/artemis/ ucbserials/ucb51110internet/1989/ucb51110442internet.pdf

Galloway, R. K., Helander, P., MacKinnon, A. L., & Brown, J. C. (2010). Thermal- isation and hard X- ray bremsstrahlung efficiency of self-interacting solar flare fast electrons. *A&A*, *520* , A72. Retrieved from https://doi.org/10.1051/ 0004-6361/201014077 doi: 10.1051/0004-6361/201014077

Giacalone, J. (2005, 5). Particle Acceleration at Shocks Moving through an Irregular Magnetic Field. *The Astrophysical Journal*, *624* (2), 765–772. Retrieved from https://doi.org/10.1086/429265 doi: 10.1086/429265

Giacalone, J., & Neugebauer, M. (2008). The Energy Spectrum of Energetic Parti- cles Downstream of Turbulent Collisionless Shocks. *The Astrophysical Journal*, *673* (1), 629–636. doi: 10.1086/524008

Gold, R. E., Krimigis, S. M., Hawkins, S. E., Haggerty, D. K., Lohr, D. A., Fiore, E., . . . Lanzerotti, L. J. (1998). Electron, proton, and alpha monitor on the advanced composition explorer spacecraft. *Space Science Reviews*, *86* (1-4), 541–562. doi: 10.1007/978-94- 011-4762-0 19

Gopalswamy, N., Xie, H., Akiyama, S., Mäkelä, P. A., & Yashiro, S. (2014). Major solar eruptions and high-energy particle events during solar cycle 24. *Earth, Planets and Space*, *66* (1), 104. Retrieved from https://doi.org/10.1186/ 1880-5981-66-104 doi: 10.1186/1880-5981-66-104

Gopalswamy, N., Yashiro, S., Akiyama, S., Mäkelä, P., Xie, H., Kaiser, M. L., . . . Bougeret, J. L. (2008). Coronal mass ejections, type ii radio bursts, and solar energetic particle events in the soho era. *Annales Geophysicae*, *26* (10), 3033– 3047. Retrieved from https://angeo.copernicus.org/articles/26/3033/ 2008/ doi: 10.5194/angeo-26-3033-2008

Harris, C. R., Millman, K. J., van der Walt, S. J., Gommers, R., Virtanen, P., Cour- napeau, D., . . . Oliphant, T. E. (2020, September). Array programming with NumPy. *Nature*, *585* (7825), 357– 362. Retrieved from https://doi.org/ 10.1038/s41586-020-2649-2 doi: 10.1038/s41586-020- 2649-2

Horne, R. B., Glauert, S. A., Meredith, N. P., Boscher, D., Maget, V., Heynderickx, D., & Pitchford, D. (2013). Space weather impacts on satellites and forecasting the earth's electron radiation belts with spacecast. *Space Weather*, *11* (4), 169–186. doi: 10.1002/swe.20023

Hunter, J. D. (2007). Matplotlib: A 2d graphics environment. *Computing in Science & Engineering*, *9* (3), 90–95. doi: 10.1109/MCSE.2007.55

Kahler, S. W. (2001). The correlation between solar energetic particle peak intensities and speeds of coronal mass ejections: Effects of ambient par- ticle intensities and energy spectra. *Journal of Geophysical Research: Space Physics*, *106* (A10), 20947-20955. Retrieved from https://agupubs .onlinelibrary.wiley.com/doi/abs/10.1029/2000JA002231 doi: https://doi.org/10.1029/2000JA002231

Kahler, S. W., & Ling, A. (2015). Dynamic sep event probability forecasts. *Space Weather*, *13* (10), 665-675. Retrieved from https://agupubs.onlinelibrary .wiley.com/doi/abs/10.1002/2015SW001222 doi: https://doi.org/10.1002/ 2015SW001222

Kahler, S. W., & Ling, A. G. (2018). Forecasting Solar Energetic Particle (SEP) events with Flare X-ray peak ratios. *Journal of Space Weather and Space Climate*, *8*, 1–10. doi: 10.1051/swsc/2018033

Kahler, S. W., White, S. M., & Ling, A. G. (2017). Forecasting E ¿ 50-MeV proton events with the proton prediction system (PPS). *Journal of space weather and space climate*, *7*, A27. doi: 10.1051/swsc/2017025

Kilpua, E. K., Lumme, E., Andreeova, K., Isavnin, A., & Koskinen, H. E. (2015). Properties and drivers of fast interplanetary shocks near the orbit of the Earth (1995-2013). *Journal of Geophysical Research: Space Physics*, *120*(6), 4112–4125. doi: 10.1002/2015JA021138

Lee, M. A. (1983). Coupled hydromagnetic wave excitation and ion acceleration at interplanetary traveling shocks. *Journal of Geophysical Research: Space Physics*, *88* (A8), 6109–6119. doi: 10.1029/JA088iA08p06109

Lemen, J. R., Title, A. M., Akin, D. J., Boerner, P. F., Chou, C., Drake, J. F., . . . Waltham, N. (2012). The Atmospheric Imaging Assembly (AIA) on the Solar Dynamics Observatory (SDO). *Solar Physics*, *275* (1-2), 17–40. doi: 10.1007/s11207-011-9776-8

Li, X., Zheng, Y., Wang, X., & Wang, L. (2020). Predicting Solar Flares Using a Novel Deep Convolutional Neural Network. *The Astrophysical Journal*, *891*(1), 10. Retrieved from http://dx.doi.org/10.3847/1538-4357/ab6d04 doi: 10.3847/1538-4357/ab6d04

Mäkelä, P., Gopalswamy, N., Akiyama, S., Xie, H., & Yashiro, S. (2011). Energetic storm particle events in coronal mass ejection-driven shocks. *Journal of Geo- physical Research: Space Physics*, *116* (8), 1–12. doi: 10.1029/2011JA016683

Marroquin, R. D., Sadykov, V., Kosovichev, A., Kitiashvili, I. N., Oria, V., Nita, G. M., . . . Ali, A. (2023, jul). Statistical study of the correlation between solar energetic particles and properties of active regions. *The Astrophysical Journal*, *952* (2), 97. Retrieved from https://dx.doi.org/10.3847/1538-4357/acdb65 doi: 10.3847/1538-4357/acdb65

Mason, G. M., Gold, R. E., Krimigis, S. M., Mazur, J. E., Andrews, G. B., Daley, K. A., . . . Walpole, P. H. (1998). The ultra-low-energy isotope spectrometer (ULEIS) for the ace spacecraft. *Space Science Reviews*, *86* (1-4), 409–448. doi: 10.1007/978-94-011-4762-0 16

Mason, G. M., & Sanderson, T. R. (1999). CIR Associated Energetic Particles in the Inner and Middle Heliosphere. *Space Science Reviews*, *89* (1), 77–90. Re- trieved from https://doi.org/10.1023/A:1005216516443 doi: 10.1023/A: 1005216516443

Maurer, R. H., Fretz, K., Angert, M. P., Bort, D. L., Goldsten, J. O., Ottman, G., . . . Bodet, D. (2017). Radiation-induced single-event effects on the van allen probes spacecraft. *IEEE Transactions on Nuclear Science*, *64* (11), 2782–2793. doi: 10.1109/TNS.2017.2754878

McComas, D. J., Bame, S. J., Barker, P., Feldman, W. C., Phillips, J. L., Riley, P., & Griffee, J. W. (1998). Solar wind electron proton alpha monitor (SWEPAM) for the advanced composition explorer. *Space Science Reviews*, *86* (1-4), 563– 612. doi: 10.1007/978-94-011-4762-0 20

Moreland, K., Dayeh, M. A., Li, G., Farahat, A., Ebert, R. W., & Desai, M. I. (2023, oct). Variability of interplanetary shock and associated energetic par- ticle properties as a function of the time window around the shock. *The Astrophysical Journal*, *956* (2), 107. Retrieved from https://dx.doi.org/ 10.3847/1538-4357/acec6c doi: 10.3847/1538-4357/acec6c

Mostafavi, P., Zank, G. P., & Webb, G. M. (2018). The Mediation of Collisionless Oblique Magnetized Shocks by Energetic Particles. *The Astrophysical Journal*, *868* (2), 120. Retrieved from http://dx.doi.org/10.3847/1538-4357/aaeb91 doi: 10.3847/1538-4357/aaeb91

Nishizuka, N., Sugiura, K., Kubo, Y., Den, M., & Ishii, M. (2018). Deep Flare Net (DeFN) Model for Solar Flare Prediction. *The Astrophysical journal*, *858* (2), 113. doi: 10.3847/1538-4357/aab9a7

Onorato, G., Di Schiavi, E., & Di Cunto, F. (2020). Understanding the ef- fects of deep space radiation on nervous system: The role of genetically tractable experimental models. *Frontiers in Physics*, *8* (October), 1–11. doi: 10.3389/fphy.2020.00362

pandas development team, T. (2020, February). pandas-dev/pandas: Pan- das. Retrieved from https://doi.org/10.5281/zenodo.3509134 doi: 10.5281/zenodo.3509134

Papaioannou, A., Sandberg, I., Anastasiadis, A., Kouloumvakos, A., Georgoulis, M. K., Tziotziou, K., . . . Hilgers, A. (2016). Solar flares, coronal mass ejections and solar energetic particle event characteristics. *Journal of Space Weather and Space Climate*, *6* . doi: 10.1051/swsc/2016035

Pesnell, W. D., Thompson, B. J., & Chamberlin, P. C. (2012). The Solar Dynam- ics Observatory (SDO). *Solar Physics*, *275* (1-2), 3–15. doi: 10.1007/s11207-011 -9841-3

Posner, A. (2007). Up to 1-hour forecasting of radiation hazards from solar ener- getic ion events with relativistic electrons. *Space Weather*, *5* (5). Retrieved from https://agupubs.onlinelibrary.wiley.com/doi/abs/10.1029/ 2006SW000268 doi: https://doi.org/10.1029/2006SW000268

Reames, D. V. (2001). Seps: Space weather hazard in interplanetary space. In *Space weather* (p. 101-107). American Geophysical Union (AGU). Retrieved from https://agupubs.onlinelibrary.wiley.com/doi/abs/10.1029/GM125p0101 doi: https://doi.org/10.1029/GM125p0101

Reames, D. V. (2013). The Two Sources of Solar Energetic Particles. *Space science reviews*, *175* (1-4), 53–92. doi: 10.1007/s11214-013-9958-9

Reames, D. V. (2017). The Abundance of Helium in the Source Plasma of Solar En- ergetic Particles. *Solar physics*, *292* (11), 1–20. doi: 10.1007/s11207-017-1173 -5

Ryan, D. F., Dominique, M., Seaton, D., Stegen, K., & White, A. (2016). Effects of flare definitions on the statistics of derived flare distributions. *Astronomy and Astrophysics*, *592* . doi: 10.1051/0004-6361/201628130

Scherrer, P. H., Bogart, R. S., Bush, R. I., Hoeksema, J. T., Kosovichev, A. G., Schou, J., . . . Team, T. M. D. I. E. (1995). The Solar Oscillations Investiga- tion - Michelson Doppler Imager. *Solar Physics*, *162* (1), 129–188. Retrieved from https://doi.org/10.1007/BF00733429 doi: 10.1007/BF00733429

Schou, J., Scherrer, P. H., Akin, D. J., Allard, B. A., Miles, J. W., Rairden, R., . . . Duvall, J., T. L. (2012, 1). Design and ground calibration of the helioseis- mic and magnetic imager (hmi) instrument on the solar dynamics observa- tory (sdo). *NASA Center for AeroSpace Information (CASI).Journal Arti- cles*. Retrieved from https://libweb.lib.utsa.edu/login?url=https:// www.proquest.com/scholarly-journals/design-ground-calibration -helioseismic-magnetic/docview/2128293240/se-2?accountid=7122 (Copyright - Copyright NASA/Langley Research Center Jan 1, 2012; Last updated - 2018-12-13)

Sellers, F. B., & Hanser, F. A. (1996). Design and calibration of the GOES-8 parti- cle sensors: the EPS and HEPAD. In E. R. Washwell (Ed.), *Goes-8 and beyond* (Vol. 2812, pp. 353–364). SPIE. Retrieved from https://doi.org/10.1117/ 12.254083 doi: 10.1117/12.254083

Shea, M. A., & Smart, D. F. (2012). Space Weather and the Ground-Level Solar Proton Events of the 23rd Solar Cycle. *Space Science Reviews*, *171* (1), 161– 188. Retrieved from https://doi.org/10.1007/s11214-012-9923-z doi: 10 .1007/s11214-012-9923-z

Smith, C. W., Heureux, J. L., & Ness, N. F. (1998). THE ACE MAGNETIC FIELDS EXPERIMENT C. W. SMITH, J. L'HEUREUX and N. F. NESS. *Space Science Reviews*, 613–632.

Stone, E. C., Cohen, C. M. S., Cook, W. R., Cummings, A. C., Gauld, B., Kecman, B., . . . Dougherty, B. L. (1998). The Solar Isotope Spectrometer for the Ad- vanced Composition Explorer. *Space science reviews*, *86* (1-4), 357–408. doi: 10.1023/A:1005027929871

Tylka, A. J., Cohen, C. M. S., Dietrich, W. F., Lee, M. A., Maclennan, C. G., Mewaldt, R. A., . . . Reames, D. V. (2005, 5). Shock Geometry, Seed Pop- ulations, and the Origin of Variable Elemental Composition at High Ener- gies in Large Gradual Solar Particle Events. *The Astrophysical Journal* , *625* (1), 474–495. Retrieved from https://doi.org/10.1086/429384 doi: 10.1086/429384

Virtanen, P., Gommers, R., Oliphant, T. E., Haberland, M., Reddy, T., Cournapeau, D., . . . SciPy 1.0 Contributors (2020). SciPy 1.0: Fundamental Algorithms for Scientific Computing in Python. *Nature Methods*, *17* , 261–272. doi: 10.1038/s41592-019-0686-2

Waskom, M. L. (2021). seaborn: statistical data visualization. *Journal of Open Source Software*, *6* (60), 3021. Retrieved from https://doi.org/10.21105/ joss.03021 doi: 10.21105/joss.03021

Xiong, B., Wan, W., Zhao, B., Yu, Y., Wei, Y., Ren, Z., & Liu, J. (2014). Re- sponse of the american equatorial and low-latitude ionosphere to the x1.5 solar flare on 13 september 2005. *Journal of Geophysical Research: Space Physics*, *119* (12), 10,336-10,347. Retrieved from https://agupubs.onlinelibrary .wiley.com/doi/abs/10.1002/2014JA020536 doi: https://doi.org/10.1002/ 2014JA020536

Yasyukevich, Y., Astafyeva, E., Padokhin, A., Ivanova, V., Syrovatskii, S., & Podlesnyi, A. (2018). The 6 september 2017 x-class solar flares and their impacts on the ionosphere, gnss, and hf radio wave propagation. *Space Weather* , *16* (8), 1013-1027. Retrieved from https://agupubs.onlinelibrary .wiley.com/doi/abs/10.1029/2018SW001932 doi: https://doi.org/10.1029/ 2018SW001932

Zank, G. P., Kryukov, I. A., Pogorelov, N. V., & Shaikh, D. (2010). The interac- tion of turbulence with shock waves. *AIP Conference Proceedings*, *1216* (April 2010), 563–567. doi: 10.1063/1.3395927